\documentclass{aa}  

\usepackage{graphicx}
\usepackage{caption}
\usepackage{txfonts}
\usepackage{float}
\usepackage[hyperindex, breaklinks, colorlinks, linkcolor=blue, citecolor=blue]{hyperref}
\usepackage{pifont}
\usepackage[switch]{lineno}
\usepackage{capt-of}
\usepackage{placeins}
\usepackage{dblfloatfix}
\usepackage{xcolor}
\usepackage{color}
\usepackage{afterpage}
\usepackage{orcidlink}
\usepackage{cuted}
\usepackage{needspace}
\usepackage{url}
\usepackage{subcaption}
\titlerunning{Intra-night optical polarization monitoring of blazars}
\authorrunning{A. Polychronakis et al.}

\begin{document}

   \title{Intra-night optical polarization monitoring of blazars}


\author{
Aristeidis Polychronakis \inst{\ref{AstroCrete2}, \ref{AstroCrete}}\thanks{\href{mailto:apolychronakis@physics.uoc.gr}{apolychronakis@physics.uoc.gr}} \orcidlink{0009-0005-7962-6296}, 
Ioannis Liodakis \inst{\ref{AstroCrete}}\thanks{\href{mailto:liodakis@ia.forth.gr}{liodakis@ia.forth.gr}} \orcidlink{0000-0001-9200-4006}, 
Anastasia Glykopoulou \inst{\ref{AstroCrete2},\ref{AstroCrete}}, 
Dmitry Blinov \inst{\ref{AstroCrete2},\ref{AstroCrete}}, 
Iv\'{a}n Agudo \inst{\ref{InstAstro_Granada}} \orcidlink{0000-0002-3777-6182}, 
Svetlana G. Jorstad \inst{\ref{BostonUni},\ref{StPetersburg}} \orcidlink{0000-0001-6158-1708}, 
Beatriz Ag\'{i}s-Gonz\'{a}lez \inst{\ref{AstroCrete}}, 
Sara Capecchiacci \inst{\ref{AstroCrete2},\ref{AstroCrete}}, 
Alberto Floris \inst{\ref{AstroCrete2},\ref{AstroCrete}}, 
Sebastian Kielhmann \inst{\ref{AstroCrete}}, 
John A. Kypriotakis \inst{\ref{AstroCrete2},\ref{AstroCrete}}, 
Dimitrios A. Langis \inst{\ref{AstroCrete2},\ref{AstroCrete}} \orcidlink{0009-0003-9365-9073}, 
Nikos Mandarakas \inst{\ref{AstroCrete2},\ref{AstroCrete},\ref{france}}, 
Karan Pal \inst{\ref{chalmers},\ref{Erlangen}}, 
Francisco J. Aceituno \inst{\ref{InstAstro_Granada}}, 
Giacomo Bonnoli \inst{\ref{InstAstro_Granada}, \ref{INAF_Merate}} \orcidlink{0000-0003-2464-9077}, 
V\'{i}ctor Casanova \inst{\ref{InstAstro_Granada}}, 
Gabriel Emery \inst{\ref{InstAstro_Granada}}, 
Juan Escudero Pedrosa \inst{\ref{InstAstro_Granada}, \ref{Harvard_Smithsonian}} \orcidlink{0000-0002-4131-655X}, 
Jorge Otero-Santos \inst{\ref{InstAstro_Granada},\ref{INFN-Padova}}, 
Daniel Morcuende \inst{\ref{InstAstro_Granada}}, 
Alfredo Sota \inst{\ref{InstAstro_Granada}} \orcidlink{0000-0002-9404-6952}, 
Vilppu Piirola \inst{\ref{UTU}} 
}

\institute{
Department of Physics, University of Crete, GR-70013 Heraklion, Greece \label{AstroCrete2} 
\and
Institute of Astrophysics, Foundation for Research and Technology-Hellas, Vasilika Vouton, GR-70013 Heraklion, Greece \label{AstroCrete} 
\and
Institute for Astrophysical Research, Boston University, 725 Commonwealth Avenue, Boston, MA 02215, USA \label{BostonUni}
\and
Saint Petersburg State University, 7/9 Universitetskaya nab., St. Petersburg, 199034 Russia \label{StPetersburg}
\and
Instituto de Astrof\'{i}sica de Andaluc\'{i}a, IAA-CSIC, Glorieta de la Astronom\'{i}a s/n, 18008 Granada, Spain \label{InstAstro_Granada}
\and
Aix Marseille Univ, CNRS, CNES, LAM, Marseille, France \label{france}
\and
Department of Space, Earth and Environment, Chalmers University of Technology, Gothenburg, Sweden\label{chalmers}
\and
Dr. Karl Remeis-Observatory and Erlangen Centre for Astroparticle Physics, Friedrich-Alexander Universit\"{a}t Erlangen-N\"{u}rnberg, Sternwartstr. 7, 96049 Bamberg, Germany\label{Erlangen}
\and
INAF Osservatorio Astronomico di Brera, Via E. Bianchi 46, 23807 Merate (LC), Italy \label{INAF_Merate}
\and
Center for Astrophysics $|$ Harvard \& Smithsonian , 60 Garden Street, Cambridge, MA 02138 USA \label{Harvard_Smithsonian}
\and
Istituto Nazionale di Fisica Nucleare, Sezione di Padova, 35131 Padova, Italy\label{INFN-Padova}
\and
Department of Physics and Astronomy, University of Turku, FI-20014, Finland \label{UTU}
}


\abstract{Blazars are known for their extreme variability across the electromagnetic spectrum. Variability at very short timescales can push the boundaries between competing models offering us much needed discriminating power. This is particularly true for polarization variability that allows us to probe particle acceleration and high-energy emission models in blazars. Here we present results from the first pilot study of intra-night optical polarization monitoring conducted using RoboPol at the Skinakas Observatory and supplemented by observations from the Calar Alto, Perkins, and Sierra Nevada observatories. Our results show that while variability patterns can widely vary between sources, variability on timescales as short as minutes is prevalent in blazar jets. The amplitude of variations are typically small, a few percent for the polarization degree and less than 20$^\circ$ for the polarization angle,  pointing to a significant contribution to the optical emission from a turbulent magnetic field component, while the overall stability of the polarization angle over time points to a preferred magnetic field orientation.}

   \keywords{Polarization -- Relativistic processes -- Galaxies: active -- BL Lacertae objects: general -- Galaxies: jets}

   \maketitle
%

\section{Introduction}

Blazars are the brightest active galactic nuclei (AGN) due to the preferential alignment of their jets towards Earth \citep{Blandford2019,Hovatta2019}. Their multiwavelength emission spans decades in energy from the lowest radio frequencies to the very high energy $\gamma$-rays. Their spectral energy distribution is typically characterized by two broad emission components (often referred to as humps) intersecting in the eV-MeV range depending on the spectral subclass. The first hump is due to synchrotron radiation while the second hump is thought to originate from either Compton scattering of relativistic electrons or emission from relativistic protons, with recent observations favoring the former over the latter \citep{Agudo2025,Liodakis2025}. The spectral subclasses in blazars are defined according to the location of the synchrotron peak into low- (LSP, $<10^{14}$Hz), intermediate -  (ISP, $10^{14}-10^{15}$Hz), and high - (HSP, $>10^{15}$Hz) synchrotron peaked sources \citep{Ajello2020}. 

The origin of the particle acceleration happening in the jets and the origin of the high energy emission (second hump) are still a matter of debate, even though a lot of progress has been made in the past few years \cite[e.g.,][]{DiGesu2023,Peirson2023,Middei2023,Marshall2023}. Polarization variability can be a powerful tool to differentiate between different models. In 2013 the RoboPol survey started at the Skinakas observatory in Crete to uncover the relation of polarization variability to $\gamma$-ray emission \citep{Blinov2015,Blinov2018,Blinov2021}. The program ran from 2013 to 2017 observing a large number of blazars with regular few day cadence. The RoboPol survey was able to triple the number of known electron vector position angle (EVPA) rotations \citep{Blinov2015,Blinov2016,Blinov2016-II} and demonstrate their connection to $\gamma$-ray flares \citep{Blinov2018}, as well as uncover the anti-correlation between polarization degree and synchrotron peak, predicting the energy stratification of the electrons \citep{Angelakis2016} that was later confirmed by X-ray polarization observations of HSPs \cite[e.g.,][]{Liodakis2022,DiGesu2022,Kouch2024}. 

\begin{table*}[h]
\caption{Summary of polarization and variability properties for the sources included in this study. The first two columns list the RoboPol ID and alternative source names. The third and fourth columns give the median polarization degree ($\Pi$) and polarization angle ($\Psi$) across all observations of each source presented here averaged in the Stokes Q-U space. The polarization angles are modified with the $\pi$ ambiguity so all the values in the column are between -90 and 90 degrees. Columns $\rm N_{tot}$ and $\rm N_\mathrm{var}$ report the total number of nights observed and the number of nights where variability was detected respectively. In the $\Pi_\mathrm{var}$ and $\Psi_\mathrm{var}$ columns, a Y indicates that variability was detected in $\Pi$ or $\Psi$. The following column shows the average cadence for each source, i.e. the median time interval between two consecutive observations, and the last two columns show the median value and its standard deviation for the 1/$\eta$ metric for $\Pi$ and $\Psi$ respectively.}

\centering
\makebox[\textwidth][c]{%
{
\begin{tabular}{c c c c c c c c c c c c}
\hline\hline
RBPL ID & Alt. ID & $\Pi$ (\%) & $\Psi$ (deg) & $\rm N_{tot}$ & $\rm{N_{var}}$ & $\rm \Pi_{var}$ & $\rm \Psi_{var}$ & Cad. (min) & $\Pi_{1/\eta}$ & $\Psi_{1/\eta}$\\
\hline
J0211+1051 & CGRaBS~J0211+1051 & 8.0 ± 2.7 & 76.8 ± 3.3 & 5 & 4 & Y & Y & 2.17 & 2 ± 1 & 1.4 ± 0.3 \\
J0217+0837 & CGRaBS~J0217+0837 & 14.0 ± 2.4 & 66.2 ± 3.1 & 7 & 4 & Y & Y & 2.67 & 0.6 ± 0.6 & 0.5 ± 0.7 \\
J0958+6533 & S4~0954+65 & 6.82 ± 0.54 & 1.8 ± 4.2 & 1 & 1 & Y & Y & 5.20 & 0.7 & 1.4  \\
J1037+5711 & GB6~J1037+5711 & 9.0 ± 2.4 & 32.3 ± 4.2 & 2 & 0 & N & N & 3.17 & 0.8 ± 0.5 & 0.44 ± 0.01  \\
J1058+5628 & TXS 1055+567 & 2.86 ± 0.65 & -0.4 ± 7.4 & 1 & 0 & N & N & 7.06 & 0.5 & 0.5  \\
J1248+5820 & TXS~1246+586 & 5.5 ± 1.0 & -77 ± 34 & 5 & 0 & N & N & 3.17 & 0.4 ± 0.1 & 0.5 ± 0.3  \\
J1512-0905 & PKS~1510-089 & 3.15 ± 0.81 & 6 ± 94 & 1 & 1 & N & Y & 2.17 & 0.6 & 3.3  \\
J1542+6129 & GB6~J1542+6129 & 3.9 ± 1.5 & 0.1 ± 4.7 & 3 & 0 & N & N & 2.17 & 0.6 ± 0.1 & 0.5 ± 0.2  \\
J1748+7005 & S5~1749+70 & 6.09 ± 0.94 & -62.0 ± 8.0 & 3 & 1 & N & Y & 1.83 & 0.58 ± 0.04 & 0.57 ± 0.06  \\
J1754+3212 & RGB~J1754+322 & 2.35 ± 0.81 & 36 ± 34 & 3 & 1 & Y & Y & 3.43 & 0.4 ± 0.3 & 0.6 ± 0.2  \\
J1800+7828 & S5~1803+78 & 5.55 ± 0.99 & 89.0 ± 7.5 & 1 & 1 & N & Y & 3.35 & 0.70 & 0.70  \\
J1806+6949 & S4~1807+69 & 8.34 ± 0.89 & 64.8 ± 1.7 & 3 & 3 & Y & Y & 1.50 & 0.59 ± 0.05 & 0.6 ± 0.1  \\
J1903+5540 & TXS~1902+556 & 6.7 ± 1.2 & 35.0 ± 4.0 & 7 & 4 & Y & Y & 3.58 & 0.67 ± 0.08 & 0.5 ± 0.1  \\
J2005+7752 & S5~2007+77 & 5.9 ± 1.6 & -75.2 ± 5.1 & 2 & 2 & Y & Y & 3.17 & 0.8 ± 0.2 & 0.6 ± 0.2  \\
J2030-0622 & TXS~2027-065 & 1.02 ± 0.36 & -44 ± 21 & 2 & 1 & N & Y & 6.83 & 0.8 ± 0.4 & 0.6 ± 0.2  \\
J2143+1743 & S3~2141+17 & 1.51 ± 0.49 & -48 ± 13 & 1 & 1 & Y & Y & 6.43 & 0.8 & 0.6  \\
J2148+0657 & 4C+06.69 & 2.2 ± 1.5 & -32 ± 71 & 7 & 4 & Y & Y & 6.82 & 0.7 ± 0.1 & 0.5 ± 0.2  \\
J2202+4216 & BL~Lacertae & 7.0 ± 4.2 & 55 ± 30 & 7 & 5 & Y & Y & 5.65 & 1 ± 1 & 1 ± 3  \\
J2232+1143 & CTA~102 & 3.2 ± 3.3 & -85 ± 83 & 3 & 2 & Y & Y & 2.68 & 0.9 ± 0.8 & 0.7 ± 0.2  \\
J2253+1608 & 3C~454.3 & 3.33 ± 0.98 & 89.4 ± 6.3 & 2 & 2 & Y & Y & 3.17 & 0.9 ± 0.2 & 0.80 ± 0.05  \\
\hline
\end{tabular}}
}
\label{table:results}
\end{table*}

However, in their analysis they found that the nominal few-day cadence of the survey was likely undersampling the light curves, resulting in masking fast EVPA rotations \citep{Kiehlmann2021}. Such rotations can be very important in differentiating between particle acceleration mechanisms (e.g., \citealp{Liodakis2024}). Intra-night variability of the total intensity has been studied in several blazars \cite[e.g.,][]{Sagar2004,Goyal2012,Bachev2012,Negi2023,Agarwal2023,McCall2024}, however, intra-night polarization variability studies have been limited to only a few selected objects \cite[e.g.,][]{Villforth2009,Fraija2017,Bachev2023,Liodakis2024}.

Building on the previous Robopol results, as well as recent findings of the Imaging X-ray Polarimetry Explorer \citep{Chen2024,Marscher2024,Capecchiacci2025}, we conducted a pilot survey in 2024 - early 2025 to uncover intra-night variable sources in polarization that could potentially show these very fast polarization angle rotations.

In Sect. \ref{sec:sample}, we discuss the sample selection and analysis. In Sect. \ref{sec:results}, we present the results from the pilot survey and a comparison with the previous RoboPol monitoring. In Sect. \ref{sec:conclusions}, we present our conclusions.


\section{Sample \& Analysis}\label{sec:sample}

Our sample is a subset of the statistically complete RoboPol survey sample of $\gamma$-ray loud blazars \citep{Pavlidou2014,Blinov2021}. From the original sample we have selected all LSP sources with a Gaia G-band magnitude brighter than 16.5. This results in twenty sources listed in Table \ref{table:results}.

The analysis of the observations was done using the semi-automatic RoboPol pipeline. Details can be found in \cite{King2014,Panopoulou2015,Blinov2021}. The observations were corrected for instrumental polarization and angle rotation using unpolarized and polarized standard stars \citep{Blinov2023}. The standard stars are observed every night and then averaged over several nights to achieve a better characterization of the instrumental polarization. The individual observing times for the sources in the sample vary based on the previous RoboPol observations. For the analysis presented here, to achieve a $>3\sigma$ detection of the polarization degree, adjacent exposures were adaptively binned to maximize signal-to-noise while maintaining a high cadence. We start by binning two adjacent data points and evaluate if the number of upper limits in the light curve is significantly reduced. If not, we repeat the procedure for three adjacent data points. The binning was applied only to individual nights and not the entire dataset for a given source. Nevertheless, this process was only applied to four sources, namely J1512-0905, J1800+7828, J2143+1743, and J2148+0657.

In addition to RoboPol, we supplement our observations with data taken between 2024-2025 as part of the Boston University BEAM-ME\footnote{\url{https://www.bu.edu/blazars/BEAM-ME.html}} program \citep{Marscher2021} and the Calar Alto \& Sierra Nevada Observatory (SNO) blazar monitoring program \citep{Agudo2012,otero2024}. BEAM-ME is using the 1.8m telescope of the Perkins Observatory (Flagstaff, AZ, USA) and the PRISM polarimeter. The Calar Alto and SNO observations are taken using CAFOS and DIPOL-1 respectively, and are analyzed with the IOP4 pipeline described in \cite{juan_escudero:2023,escudero2024}.  The additional observations were typically within a few days from the RoboPol observations. For J1058+5628 we were able to obtain only one night of observations using the Perkins telescope. The polarization degree has been debiased following \cite{Blinov2021}. The contribution of the interstellar polarization for the sources in our sample is typically $<1\%$, and any depolarization from the host galaxy for LSP sources is typically negligible. Regardless, since we are interested in identifying variability, our results should not be affected by either effect. All the raw measurements of the Stokes parameters from  our pilot survey without any quality cuts, corrections or binning are available in the Harvard dataverse\footnote{\cite{DVN/G0FYM0_2025}, \url{https://dataverse.harvard.edu/dataset.xhtml?persistentId=doi:10.7910/DVN/G0FYM0}}. 

In total we obtained 65 nights of observations for the entire sample (see Table \ref{table:results}) with varying length due to observing conditions and seasonal night-length variations. We were able to achieve a median cadence of a few minutes, typically $<7$ minutes, with a minimum of 1.5 minutes (Table \ref{table:results}).

To assess whether the sources exhibit intra-night variability, we need to define a variability metric. We attempted a simple chi-square approach of comparing the observed data for each night against a constant value, however, such an approach suffers in two ways. First, it is highly sensitive to outliers, and second, it is not trivial to define an appropriate threshold to differentiate variable from non-variable light curves. For that reason, we partially adopted the methodology proposed by \citet{SOKOLOVSKY2017}, who compared the performance of various variability detection techniques in photometric time series. We chose to use one of their recommendations: the $1/\eta$ statistic.

\begin{figure}[H]
    \centering  
    \begin{subfigure}{0.45\textwidth}  
        \centering  
        \includegraphics[width=\textwidth, height=0.4\textheight, keepaspectratio]{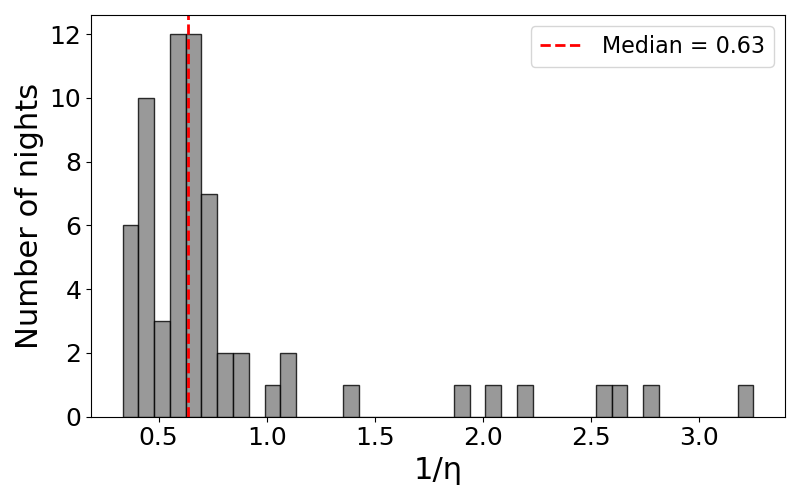}
        \caption{Polarization Degree ($\Pi$).}
        \label{DCF_res}
    \end{subfigure}
    
    \vspace{0.3cm}  
    
    \begin{subfigure}{0.45\textwidth}  
        \centering  
        \includegraphics[width=\textwidth, height=0.4\textheight, keepaspectratio]{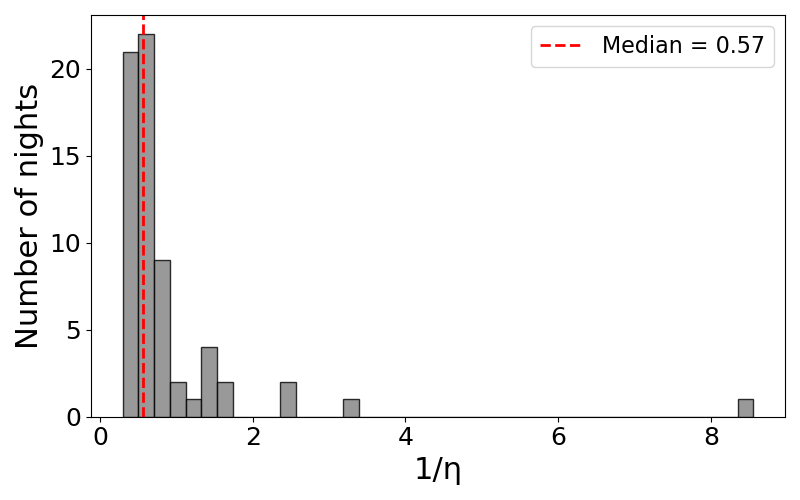}
        \caption{Polarization Angle ($\Psi$).}
        \label{ST_res}
    \end{subfigure}

    \caption{Distributions of the $1/\eta$ metric for the polarization degree ($\Pi$, upper panel) and the polarization angle ($\Psi$, lower panel), across all 65 observing nights. The vertical red dashed line in each panel indicates the median of the distribution, which also serves as the threshold separating variable from non-variable nights.}
    \label{histogramms}
\end{figure}

This index measures the degree of correlation or smoothness in a time series. It is defined as the inverse of the von Neumann ratio,
\begin{equation}
\rm{\eta = \frac{\delta^2}{\sigma^2} = 
\frac{
\displaystyle\sum_{i=1}^{N-1} (x_{i+1} - x_i)^2 / (N - 1)
}{
\displaystyle\sum_{i=1}^{N} (x_i - \bar{x})^2 / (N - 1)
}},
\end{equation}
where $\rm x_i$ is the measured value at time index i, $\rm \bar{x}$ is the mean of all $\rm x_i$, N is the total number of values, $\delta^2$ is the mean squared successive difference, and $\sigma^2$ is the sample variance. To define a threshold for the $1/\eta$ metric and identify nights as variable when exceeding this value, we adopted the median of each distribution, computed separately for $\Pi$ and $\Psi$ (see Fig. \ref{histogramms}). Selecting to use the agnostic median value of the individual  $\Pi$ (median $1/\eta=$0.63) and $\Psi$  (median $1/\eta=$0.57) distributions was motivated by the fact that any choice of threshold could have an impact on the results. The median was also chosen because the distributions deviate from a Gaussian shape, making the mean of the distributions a less representative metric, while the mode of the distributions lie very close to their medians. A higher value of $1/\eta$ indicates smoother, more coherent evolution over time—characteristic of real variability. In contrast, purely random scatter leads to values of 1/$\eta$ close to or below the median $\Pi$ and $\Psi$.

Given the stochastic nature of blazar variability, we complement the $1/\eta$ criterion, by computing for each night both the standard deviation of the measurements and their median uncertainty. If a night exceeds the $1/\eta$ threshold, we assess whether the amplitude of its variation is significant by comparing the standard deviation with the median uncertainty within the night. If the standard deviation is larger than the median uncertainty, it would suggest that the observed excursions are likely intrinsic to the source and not due to the signal-to-noise of the observations.  Using this additional check, we are able to evaluate the overall behavior of the fluctuations —independent of temporal structure— and remain robust against outliers.

By combining the $1/\eta$ metric, which is sensitive to temporal correlations, with the standard deviation relative to the measurement uncertainties, which reflects the amplitude of variations, we obtain an overall more robust criterion for identifying genuine variability. 

\begin{figure*}[htbp]
    \centering
    \includegraphics[width=0.9\textwidth, keepaspectratio]{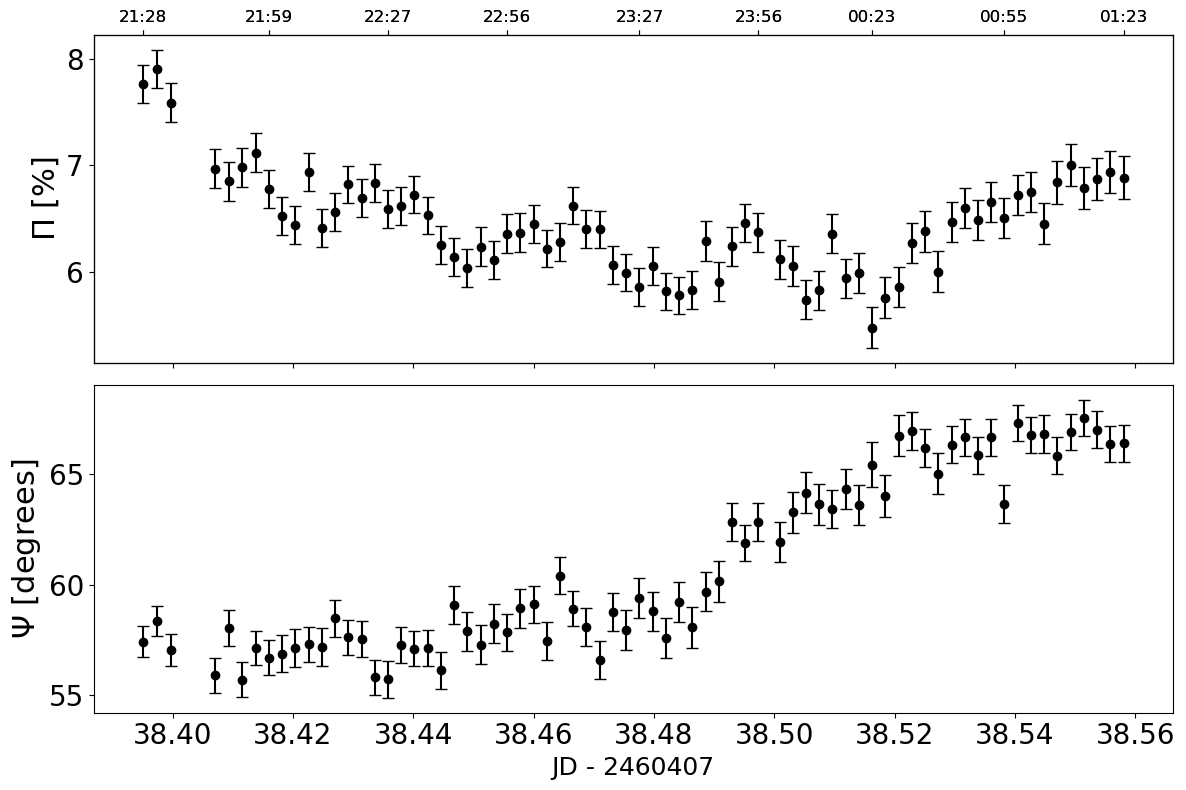}
    \caption{Intra-night RoboPol observations of BL Lac (example of a source showing polarization variability). The top panel shows the polarization degree, and the bottom panel the polarization angle. The observations have an average cadence of 3.31 minutes.}
    \label{bllac}
\end{figure*}

\begin{figure*}[htbp]
    \centering
    \includegraphics[width=0.9\textwidth, keepaspectratio]{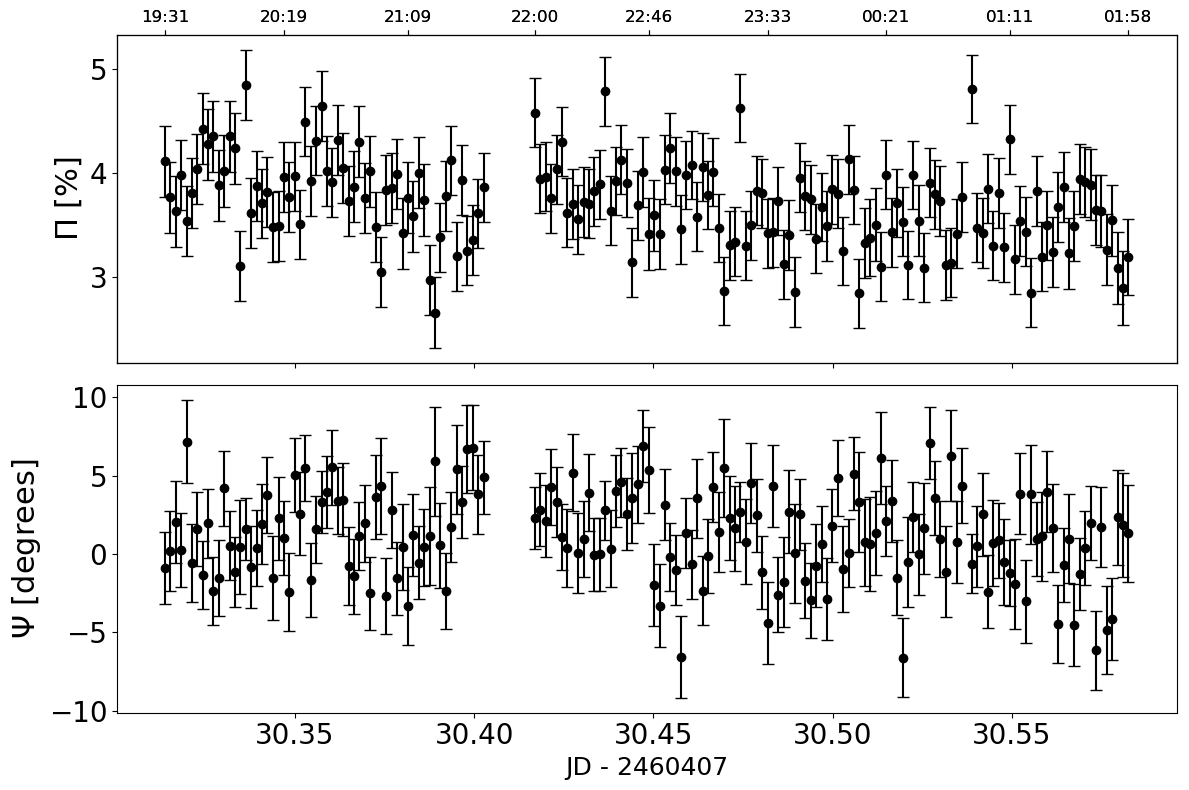}
    \caption{Intra-night RoboPol observations of J1542+6129 (example of a source without polarization variability). The top panel shows the polarization degree, and the bottom panel the polarization angle. The observations have an average cadence of 2.29 minutes.}
    \label{source}
\end{figure*}

\begin{figure*}[htbp]
    \centering
    \includegraphics[width=0.9\textwidth, keepaspectratio]{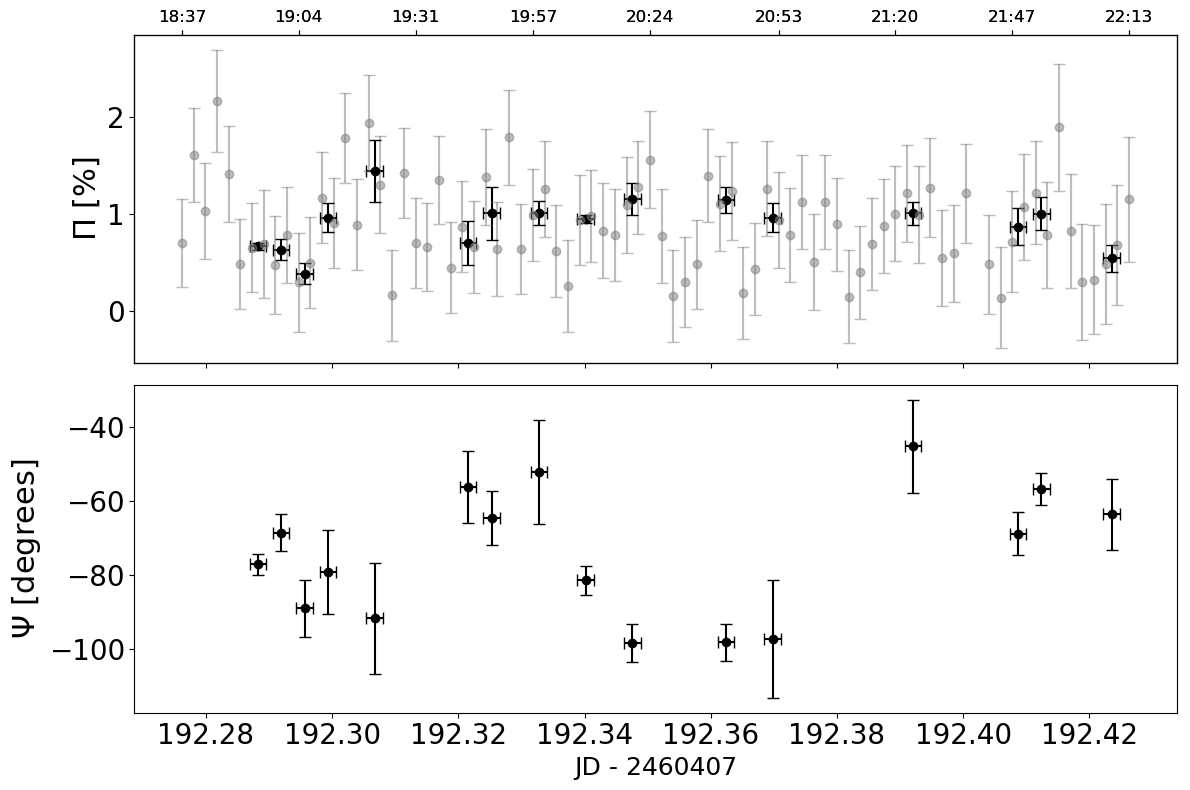}
    \caption{Intra-night RoboPol observations of J2148+0657 (example of a source with binned observations; two consecutive images combined). The grey points show the unbinned observations. The top panel shows the polarization degree, and the bottom panel the polarization angle. The upper limits have been omitted for clarity. The observations have an average cadence of 12.99 minutes.}
    \label{binned_source}
\end{figure*}

In Fig. \ref{bllac} we present the results for BL Lacertae (J2202+4216), which serves as an example of a source exhibiting clear intra-night variability. In contrast, Fig. \ref{source} shows the results for J1542+6129, a case with no significant intra-night variability. Fig. \ref{binned_source} illustrates a source whose polarization degree ($\Pi$) remains very close to zero, resulting in many non-detections ($\Pi<3\cdot\delta\Pi$) throughout the night. For this target, we binned the data in pairs (combining two consecutive images) and retained only the detection points. It is important to note that this binning procedure was performed on the Stokes Q and Stokes U space, and then the $\Pi$ and $\Psi$  were recomputed according to the following standard equations,

\begin{equation}
    \rm \Pi = \sqrt{\rm Q^2 + U^2},
\end{equation}

\begin{equation}
    \rm{\Psi = \frac{1}{2} \arctan\left( \frac{U}{Q} \right).}
\end{equation}

\section{Results}\label{sec:results}

Among the 65 observing nights in total, 37 show evidence for variability either in the polarization degree or angle. Table \ref{table:results} presents our results for all sources with columns indicating the total number of observations for each source and the number of nights we find evidence for polarization variability. The reported values of $\Pi$ and $\Psi$ in the Table correspond to the median (and standard deviation) over all available measurements (averaged in Stokes Q-U space), rather than intra-night values which are available on the Harvard dataverse repository. Overall, we find a median $\Pi\approx5\%$ with a minimum of $\approx1\%$ and a maximum of $\approx14\%$, consistent with previous RoboPol observations \citep{Angelakis2016}. Intra-night variable blazars exhibit a variety of variability patterns, from smooth continuous changes throughout a single night (e.g., Fig. \ref{bllac}), to more abrupt stochastic variations. 

Given that our sample is statistically well defined and the monitoring was performed blindly (i.e., without selecting particular nights), 56.9\% (37 out of 65 nights) of all blazar observations should show intra-night variability of the polarization parameters in a magnitude limited sample. Out of the twenty sources in our sample, only four did not show variability either in $\Pi$ or $\Psi$. However, three out of the four sources show low $\Pi<5\%$ which could prevent us from uncovering variability patterns due to signal-to-noise limitations. Nevertheless, that would suggest that 80\% of low synchrotron peaked blazars in a blind survey show intra-night variability. Moreover, seven sources (35\% of the sample) show variability in every observing night, while the overall duty cycle is 61.9\% (average $\rm N_{var}/N$  from all the sources in the sample -- see Table \ref{table:results}) . The amplitude of the variations is typically $\Delta\Pi<5\%$ and  $\Delta\Psi<20^\circ$. We do not find strong evidence for a preference of variability in $\Pi$ or $\Psi$.

\subsection{Comparison with archival observations} \label{sec:comparison}

\begin{figure*}[htbp]
    \centering
    \includegraphics[width=\textwidth, keepaspectratio]{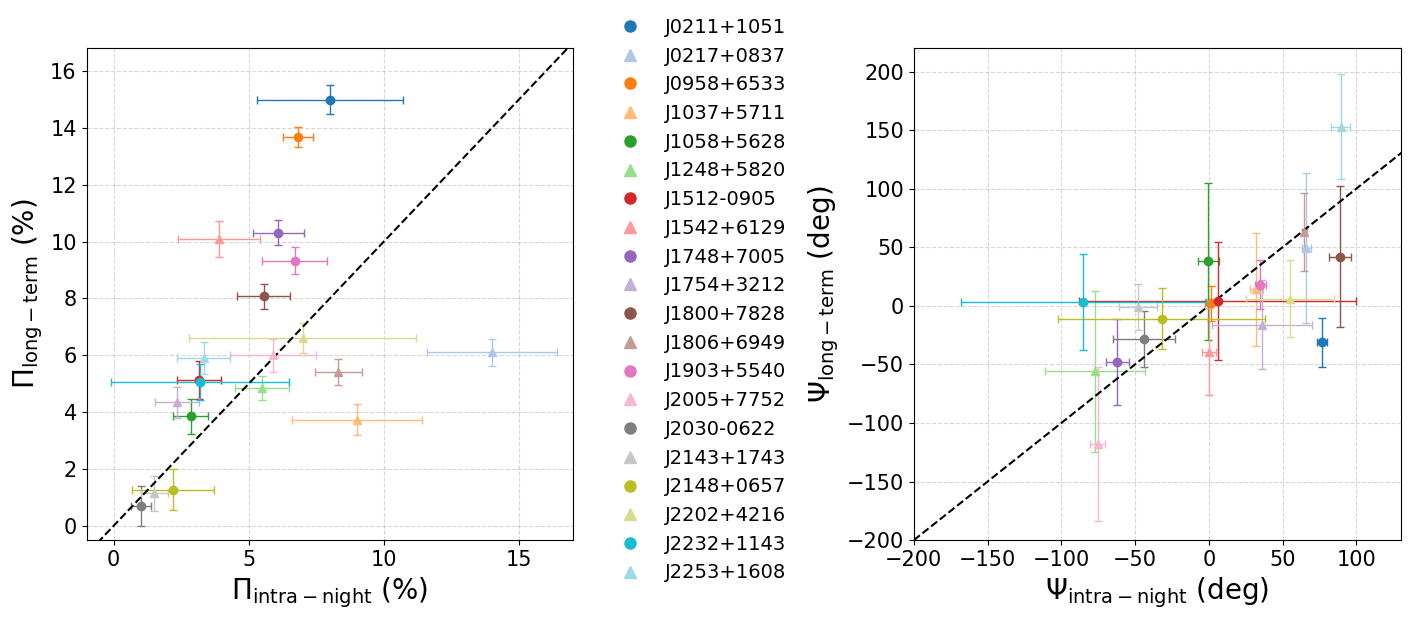}
    \caption{Comparison of our results (intra-night monitoring) with those of the previous survey (long-term monitoring; \citealt{Blinov2021}). The left panel shows the polarization degree ($\Pi$), and the right panel the polarization angle ($\Psi$). Different colors indicate different sources. The black dashed line on each panel marks equality ($\rm{y = x}$)  and illustrates the agreement between the long-term and intra-night monitoring observations.}
    \label{comparison}
\end{figure*}

To compare our results with archival data, we used measurements from the long-term monitoring survey conducted at the Skinakas Observatory from 2013 to 2017 \citep{Blinov2021}. We calculate the median and standard deviation of the Stokes parameters for each source which we then convert to $\Pi$ and $\Psi$. Fig.~\ref{comparison} presents this comparison for both $\Pi$ and $\Psi$. Our intra-night results are shown along the x-axes, while the archival values are plotted on the y-axes. The black dashed line in each panel represents equality. We find an overall good agreement between datasets, even if those are taken in completely different times and timescales. Intra-night $\Pi$ shows the largest departures from the long-term behavior with sources becoming both more and less polarized. On the other hand, the intra-night $\Psi$ shows an almost perfect agreement with the long-term observations.

%

\section{Conclusions} \label{sec:conclusions}
We presented a pilot study of twenty blazars selected from the RoboPol sample  with strict statistical criteria. The sources were observed over a total of 65 nights from 2024 until early 2025. Our dataset is further supplemented by observations from Calar Alto, Perkins, and the Sierra Nevada monitoring programs. We find that the majority of our observations show intra-night variability in either $\Pi$ or $\Psi$, with no strong evidence for a preference in either $\Pi$ or $\Psi$. Only four sources did not show variability, however the observations for three out of the four were taken in low $\Pi$ states. Future observations of these sources have the potential to reveal variability if taken in higher polarization states. For the other sources, the duty cycle for detecting intra-night variability in a blind survey has a median of 61.9\%, while the amplitude of the observed variability is small, with changes typically no more than 5\% in the polarization degree and 20 degrees in the polarization angle. 

Comparing our campaign with the long-term observations taken by RoboPol between 2013-2017 \citep{Blinov2021}, we find that the polarization degree has changed in a number of sources, showing both lower or higher polarization. Interestingly, the intra-night polarization angle appears to be consistent with the long-term behavior. Models of magnetic field variability in blazars usually invoke either shocks moving inside the jet \citep[e.g.,][]{Marscher1985,Marscher2008,Marscher2010,Liodakis2020}, magnetic reconnection \citep[e.g.,][]{Hosking2020,Zhang2020}, kink instabilities \cite[e.g.,][]{Dong2020,Jorstad2022}, or turbulence \citep[e.g.,][]{Marscher2014,Peirson2018}. The long-term stability of the polarization angle indicates the presence of a persistent magnetic field orientation. This could be either due to the presence of a recollimation shock or the presence of a large scale helical magnetic field permeating the jet \cite[e.g.,][]{Clausen-Brown2011,Hovatta2012,Paraschos2024}. On the other hand, the minute-timescale low amplitude variability, as well as the much lower observed polarization degree compared to the theoretical maximum for synchrotron radiation ($\sim70\%$), indicates the presence of a significant turbulent magnetic field component. Therefore, models that can accommodate both ordered and disordered magnetic field components \citep[e.g.,][]{Marscher2014} are necessary to account for the observed phenomenology. 

Beyond this pilot study, since early 2025  we have been performing additional high-cadence intra-night observations of blazars with the highest variability duty cycle using RoboPol, aiming to detect the predicted intra-night polarization angle rotations. These observations will be discussed in an upcoming publication.

\begin{acknowledgements}
We thank the anonymous referee for useful comments that helped improve the paper. The study was funded by the European Union ERC-2022-STG - BOOTES - 101076343. Views and opinions expressed are however those of the author(s) only and do not necessarily reflect those of the European Union or the European Research Council Executive Agency. Neither the European Union nor the granting authority can be held responsible for them. The Perkins Telescope Observatory, located in Flagstaff, AZ, USA, is owned and operated by Boston University. Observations at the Perkins telescope were supported by NASA Fermi Guest Investigator grant 80NSSC23K1507. Some of the data are based on observations collected at the Centro Astron\'{o}mico Hispano en Andalucía (CAHA), operated jointly by Junta de Andaluc\'{i}a and Consejo Superior de Investigaciones Cient\'{i}ficas (IAA-CSIC). Some of the data are based on observations collected at the Observatorio de Sierra Nevada; which is owned and operated by the Instituto de Astrof\'isica de Andaluc\'ia (IAA-CSIC). The IAA-CSIC co-authors acknowledge financial support from the Spanish "Ministerio de Ciencia e Innovaci\'{o}n" (MCIN/AEI/ 10.13039/501100011033) through the Center of Excellence Severo Ochoa award for the Instituto de Astrof\'{i}isica de Andaluc\'{i}a-CSIC (CEX2021-001131-S), and through grants PID2019-107847RB-C44 and PID2022-139117NB-C44. D.B. acknowledges support from the European Research Council (ERC) under the Horizon ERC Grants 2021 programme under grant agreement No. 101040021.
\end{acknowledgements}

\bibliographystyle{aa}  
\bibliography{bibliography}  

\clearpage

\end{document}